\documentclass[a4paper,11pt]{apa}

\usepackage{color,graphicx}
\usepackage{url,lineno,microtype}
\usepackage[onehalfspacing]{setspace}

\usepackage{amsmath}
\usepackage[T1]{fontenc}
\usepackage{CJK}
\usepackage{amssymb}
\usepackage{geometry} 
\newenvironment{chinois}{\begin{CJK}{UTF8}{gkai}}{\end{CJK}}

\begin{document}

\title{Structural Stability of Lexical Semantic Spaces: Nouns in Chinese and French} 
\rightheader{Structural Stability of Lexical Semantic Spaces}

\sixauthors{Sabine Ploux\footnote{Corresponding author. Fax: +33 4 37 91 12 10. Tel: +33 4 37 91 12 53.
E-mail address: sploux@isc.cnrs.fr. }}{Rui Wang}{ZhengFeng Zhong}{Hai Zhao}{Yang Xin}{Bao-Liang Lu}
\sixaffiliations{Institut des Sciences Cognitives Marc Jeannerod, CNRS/University of Lyon I, Bron, France}
{Center for Brain-Like Computing and Machine Intelligence, Department of Computer Science and Engineering, Shanghai Jiao Tong University, Shanghai, China\\Institut des Sciences Cognitives Marc Jeannerod, CNRS/University of Lyon I, Bron, France}
{East China Normal University, Shanghai, China\\Institut des Sciences Cognitives Marc Jeannerod, CNRS/University of Lyon I, Bron, France\\Universit\'e libre de Bruxelles, Bruxelles, Belgium}
{Center for Brain-Like Computing and Machine Intelligence, Department of Computer Science and Engineering, Shanghai Jiao Tong University, Shanghai, China}
{Center for Brain-Like Computing and Machine Intelligence, Department of Computer Science and Engineering, Shanghai Jiao Tong University, Shanghai, China}
{Center for Brain-Like Computing and Machine Intelligence, Department of Computer Science and Engineering, Shanghai Jiao Tong University, Shanghai, China}

\date{}

\abstract{\vspace*{\fill}\section*{Abstract}  

Many studies in the neurosciences have dealt with the semantic processing of words or categories, but few have looked into the semantic organization of the lexicon thought as a system. The present study was designed to try to move towards this goal, using both electrophysiological and corpus-based data, and to compare two languages from different families: French and Mandarin Chinese.

We conducted an EEG-based semantic-decision experiment using 240 words from eight categories (\emph{clothing}, \emph{parts of a house}, \emph{tools}, \emph{vehicles}, \emph{fruits/vegetables}, \emph{animals}, \emph{body parts}, and \emph{people}) as the material. A data-analysis method (correspondence analysis) commonly used in computational linguistics was applied to the electrophysiological signals.

The present cross-language comparison indicated stability for the following aspects of the languages' lexical semantic organizations: (1) the living/nonliving distinction, which showed up as a main factor for both languages; (2) greater dispersion of the living categories as compared to the nonliving ones; (3) prototypicality of the \emph{animals} category within the living categories, and with respect to the living/nonliving distinction; and (4) the existence of a person-centered reference gradient. Our electrophysiological analysis indicated stability of the networks at play in each of these processes. Stability was also observed in the data taken from word usage in the languages (synonyms and associated words obtained from textual corpora).

\vspace*{5cm}

Keywords: mental lexicon, semantics, ERP-derived semantic space, NLP-derived semantic space, Cross-linguistic comparison}

\maketitle

\section{}
Many studies in the neurosciences have dealt with the semantic processing of words or categories, but few have looked into the organization of the lexicon and its underlying concepts thought as a system (see \cite{huth2016natural} for a monolingual study).
This question is crucial however, for it is relevant to various still-largely-unexplored domains, including the construction of the lexicon, its organizational stability and/or variability, and its degradation in degenerative pathologies such as primary progressive aphasia (PPA)  \cite{hurley2009electrophysiology}.

To add to the current debate on the contingencies of word learning and word representations in the different sensorimotor modalities \cite{caramazza2014embodied,cappa2012cortex,vigliocco2009toward}, we can wonder about whether there is an organizing principle for structuring and building up the lexicon, and about how new words are added in relation to each other (i.e., their mutual positioning). And reciprocally, whenever the processing of word meaning deteriorates, it is important to understand the underlying principle: Is the degradation distributed in an arbitrary way or does it follow a plan linked to the overall structure of the lexicon? The latter possibility was supported by a study showing that lexical semantic degradation starts on single entities and then moves on to more general categories \cite{laisney2011zebra}, and also by a study in which PPA patients (especially those with semantic dementia) exhibited a poor ability to choose words that shared the same semantic field in a picture-naming task 
 \cite{vandenberghe2005paradoxical}. A final question concerns how stable this organization is and how it varies across individuals, societies, and linguistic systems. 

 The present comparative study was designed to address this last point: the organization of the lexicon in different languages. The question of cross-language similarities and differences has been widely studied in linguistics and linguistic modeling. The linguistic project Eurowordnet \cite{vossen1998}, for example, attempted to construct an \emph{a priori} organization of the lexicon that would fit the hierarchical structure of concepts and also be language-independent. The project ran into numerous obstacles stemming from difficulty in matching concepts across languages, a problem that counters the classic assumption that the conceptual level is language-independent and can capture how every language "breaks down" reality into words. This difficulty shows up locally in word semantics, for there are many words in a given language that have no direct equivalent in another language. The same holds true of the concepts underlying the lexicon, as noted by P. Vossen \cite{vossen1998}, who illustrates with the Dutch verb \emph{klunen} (\emph{to walk on skates}) that has no English counterpart. The process of word borrowing -- e.g. the Chinese words \begin{chinois}阴\end{chinois} (\emph{y\=\i n}) and  \begin{chinois}阳\end{chinois} (\emph{y\'ang}) 
were adopted by Western languages precisely because they had no lexical or even conceptual counterparts -- shows that variability is not a purely lexical phenomenon but also applies to what several authors like Miller \cite{miller} have called lexical concepts (named categories)\footnote{Note that these phenomena of lack of  lexical concept equivalent of one language to another are not only present when it comes to words denoting abstract concepts or verbs but also exist for concrete notions of the same type as those selected from our study. For example,  kinship ties in Chinese are much more precise and differentiated than in French or in English: e.g. \begin{chinois}伯婆\end{chinois} (\emph{b\'op\'o}) refers to the wife of the elder brother of the paternal grandfather.}.

 However, while the question of variability versus stability has been addressed from the local standpoint of word semantics, we know of no studies on word semantics that have looked at the lexicon's overall semantic organization, generally assumed to be uniform and hierarchical because it has a common conceptual base that is constrained and determined by our humanness.

In an earlier study \cite{ploux2012}, we applied a data analysis to ERP signals to find the main axes that organize the neurophysiological and semantic processing of 240\footnote{240 nouns do not cover all the nouns of a language. We plan to expand the set of words in further experiments.} French words.
We were interested in nouns referring to objects/persons categories that are often studied in the literature (e.g., \cite{vigliocco2007semantic}).
 In the present study, we apply the same protocol to words in Mandarin Chinese presented to native speakers of Chinese. By comparing two languages from families as different as French (an Indo-European Romance language) and Mandarin Chinese (in the Sino-Tibetan family of languages), we attempt to uncover not only differences but also similarities in their lexical organizations, from the joint viewpoint of neurophysiology and computational linguistics.

\section{Material \& Methods}
\subsection{Stimuli and Procedure}
The experiment conducted here used a semantic-decision task. The words presented to participants were nouns (written in simplified Chinese characters) belonging to eight categories: four containing nouns referring to living entities and four containing nouns referring to nonliving entities. Each category included 30 nouns, making for a total of 240. The nonliving stimuli were chosen from the following categories: clothing (e.g. \emph{dress}), part of a house (e.g. \emph{kitchen}, \emph{staircase}), tools (e.g. \emph{hammer}), and vehicles (e.g. \emph{airplane}); the living stimuli were chosen from fruits/vegetables (e.g. \emph{lemon}), animals (e.g. \emph{cow}), body parts (e.g. \emph{neck}), and people (e.g. \emph{brother}, \emph{novelist}). The lexical frequencies were controlled in the whole set of words, and also in each of the eight categories. Lexical frequencies were calculated from a large Chinese corpus (368 million characters) taken from popular newspapers in China between 1980 to 1998 ($\text{mean}(log_{10}) = 2.5$, $\text{std} = 0.92$). The mean number of characters per word was $2.21$ ($\text{std} = 0.69$). 

The participants had to state whether the words presented referred to biological or non-biological entities. This terminology seemed preferable to living versus nonliving things because it is less semantically ambiguous (Is a piece of fruit a living thing or not?). Participants were seated $\text{60 cm}$ from the screen and had to click on the right (or left) button on a box placed in front of them if they thought the displayed word fell into the biological category, and click on the opposite side (right or left) if they thought the word fell into the non-biological category. The clicking sides for the biological and non-biological categories were randomly assigned to each participant. The experiment included an initial training phase during which words from each category were presented. None of the training words belonged to the material used during the signal-recording phase. In the experiment proper, the stimuli were presented in three blocks, each lasting an average of about eight minutes. In a given block, all $240$ words were displayed once, in random order. The display timing was as follows: a mid-screen fixation cross was displayed for $\text{250 ms}$, followed by a word for $\text{800 ms}$, then a fixation cross until the participant responded. Participants were instructed to respond only after the second fixation cross appeared. A black screen lasting $\text{500 ms}$ followed the offset of the second fixation cross. Participants were asked not to blink their eyes until the end of a trial.

\subsection{Participants}
Eighteen volunteer students (9 females, 9 males, mean age $21.7$ years) at Shanghai JiaoTong University participated in the experiment. All were right-handed. None had any known neurological disorders or were taking any medication. They all reported having normal or corrected-to-normal vision. The study was approved by the Ethical Committee CPP Sud-Est II and all participants gave their written informed consent.

\subsection{EEG Recording}

The EEG signals were recorded by a Neuroscan Quik-cap device connected to 64 electrodes whose impedances were kept under $\text{10 k$\Omega$}$. 
The electrodes were positioned according to the 10/20 system. The sampling frequency was $\text{1000 Hz}$. The EEG data were pre-analyzed with Fieldtrip-matlab software \cite{oostenveld2010fieldtrip}. Segmentation windows started $\text{250 ms}$ before and ended $\text{750 ms}$ after the appearance of the word. Baseline correction using a $(\text{-250},\text{0}) ms$ window before stimulus onset and various filters (high-pass: $\text{1 Hz}$; low-pass: $\text{60 Hz}$, notch filter: $\text{50 Hz}$ ) was applied. The Fieldtrip \emph{ft\_rejectvisual} function was used for segment rejection and bad-electrode detection. Bad electrodes were repaired using triangulation on the cap surface and spline functions (Fieldtrip \emph{ft\_channelrepair}). 
EEG channels were re-referenced offline to the average reference, including all electrodes except the mastoid electrodes (M1, M2), CB1, CB2, and the electro-oculograms (EOGs)\footnote{This procedure was justified by the uniform distribution of the electrodes. Note, however, that since the calculation of the reference value and the data analysis method used were linear, this procedure did not cause a distorting effect. }. The mean of the segments was then calculated for each of the eight conditions and for each participant, along with the grand mean per condition for all participants pooled.

\subsection{Data Analysis}
A correspondence analysis (CA) \cite{benzecri} and the subsequent analyses were performed using FCarte \cite{Fcarte}, a dedicated software platform written in Matlab\circledR ~(source code is available upon request to the corresponding author). This type of analysis calculates data structures by detecting the principal axes of a cluster of points constructed from a data table. CA was used here to determine and order all interdependencies among the experimental conditions and the spatiotemporal networks of electrophysiological activity. Since lexical access was estimated to occur within the first $\text{500 ms}$ after word onset \cite{Friederici2002towards}, our analysis focused on a time window between $\text{100 ms}$ and $\text{500 ms}$ after word onset. 
The CA was performed on the matrix denoted $M_{100-500}$, which has eight rows, one per condition (i.e., category). The columns correspond to a sampling of the amplitude of the grand means of the ERP signals between $\text{100 ms}$ and $\text{500 ms}$ after stimulus onset, on the set of 60 electrodes (after removal of M1, M2, CB1, and CB2). Wavelet decimation was used to reduce the size of the matrix\footnote{The signals were smoothed in advance by applying the \emph{wavedec} and \emph{wrcoef} functions of Matlab. The \emph{wavedec} function processes the signal's wavelets. The family of wavelets retained was daubechies ($db1$) on 3 levels, $[C,L] = wavedec(signal,3,db1)$. The function \emph{wrcoef} uses the coefficients calculated by \emph{wavedec} to reconstruct the smoothed signal. The signal was reconstructed at level $3$.} while avoiding sampling noise. The sampling period was calculated automatically by applying this procedure; it was equal to $7 ms$ in the present study. Due to a large deviation of more than two standard deviations from the mean of the results, we removed one participant  from the analysis. 
Purely statistical tests were done in Statistica\circledR.

\section{Results}
We begin by presenting the CA results for the first two axes, in descending order of the percentage of explained inertia  (Figure \ref{fig:01}). Next we perform statistical tests on the topology of the axes (see Figure \ref{fig:02} for the overall map topology (Axes 1 and 2)). Then we compare these two axes to the results obtained for French \cite{ploux2012}. Lastly, we present the spatiotemporal networks underlying the organization specified by the axes.

\subsection{Analysis of the Axes}
\begin{figure}[!t]
\includegraphics[width=18cm]{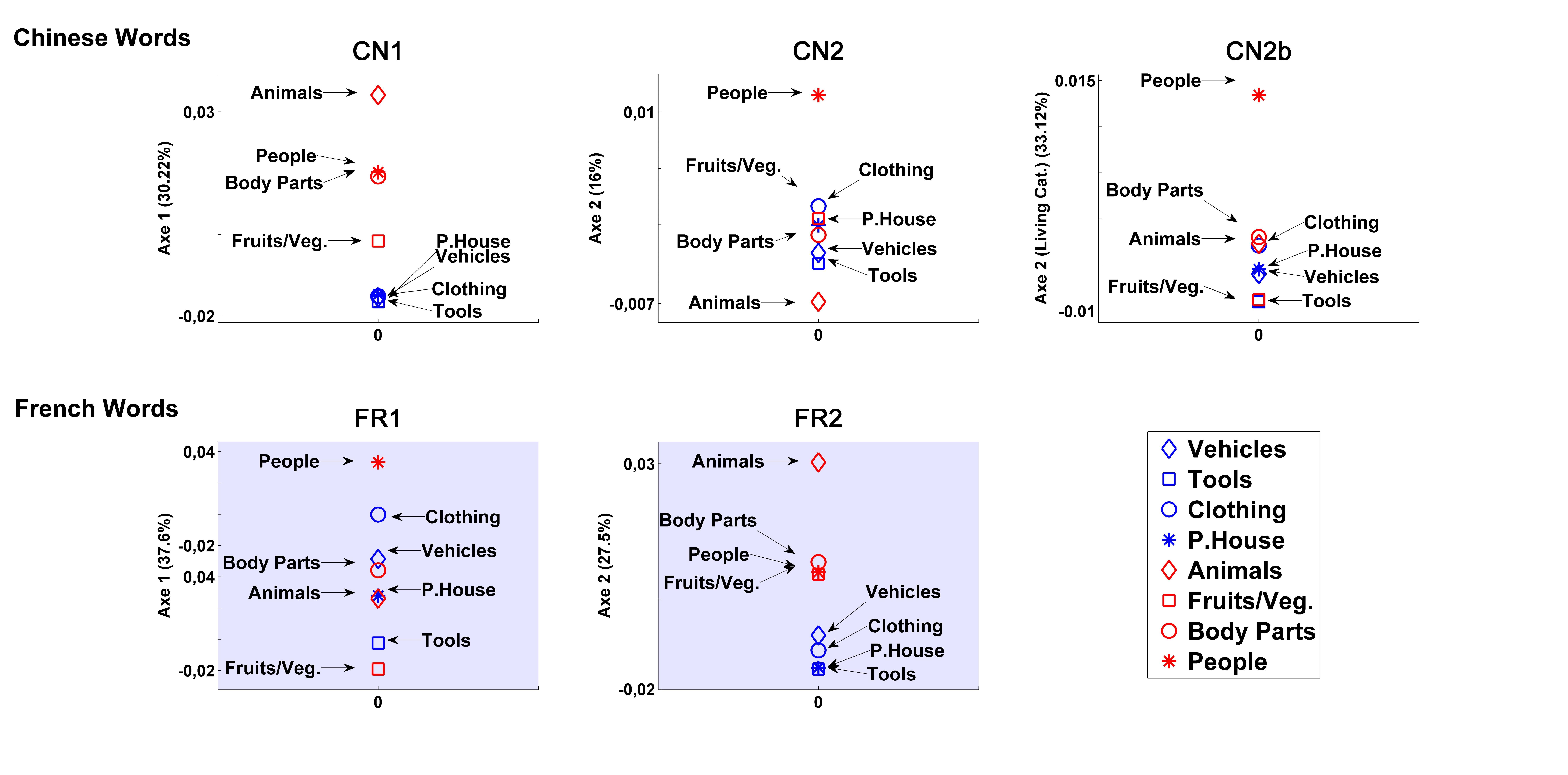}\\
 \textbf{\refstepcounter{figure}\label{fig:01} Figure 1.}{Results of the CA. Upper left (CN1) and upper middle (CN2): first two axes for Chinese words. Upper right (CN2b): second axis of living categories, with projections of the nonliving categories. Lower (FR1 and FR2) (taken from [Ploux et al., 2012]): first two axes for French words. The ordinates are the coordinates of the categories on the axis. The percentage of inertia explained by each axis in the analysis is shown in parentheses.}
\end{figure}
\begin{figure}[!h]
\includegraphics[width=18cm]{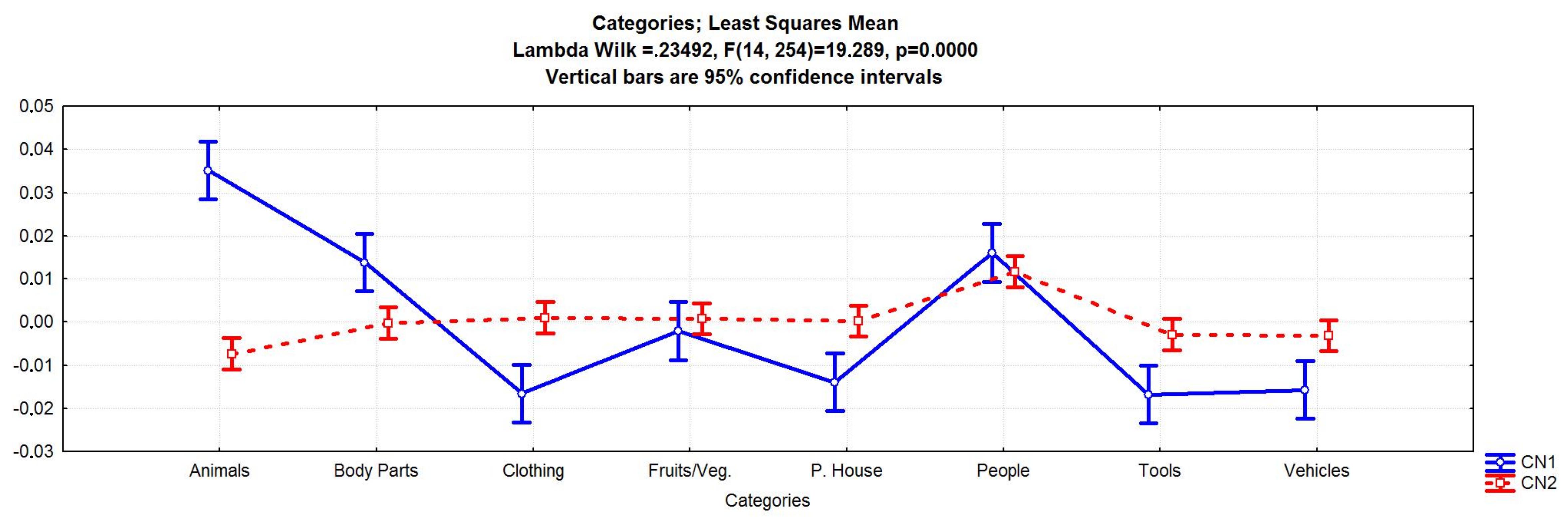}\\
\textbf{\refstepcounter{figure}\label{fig:02} Figure 2.}{ Repeated-measures ANOVA on categories. The dependent variables are the
coordinates on the two axes CN1 and CN2.}
\end{figure}

\paragraph{Principal axis: distinction between living and nonliving.}
Axis 1 (denoted CN1) of the CA (percentage of inertia 30.22\%) indicated (1) a distinction between all living categories (red in Figure \ref{fig:01}) and all nonliving categories (blue) (Anova\footnote{The statistical tests were based on projections onto the axes of the ERP means in each category, for all subjects.}  ($F(7, 112) = 29, p < .0001$, Duncan post-hoc test, Table 1), (2) more dispersion for living categories than for nonliving ones ($F(2,15) = 52, p < .001$), and (3) a topology for this axis analogous to that of axis 2 for French categories (correlation coefficient $corr(CN1, FR2) = 0.96, p < .0005$). As in French, the \emph{animals} category is the most distinct on this axis ($p(animals, other categories) < .001$). All of the living categories differ from each other except \emph{people} and \emph{body parts}.

\begin{table}[!t]
\textbf{\refstepcounter{table}\label{Tab:01} Table 1.}{ Duncan post-hoc Test ($*10^{-3}$)  on the coordinates, for all subjects (repeated-measures ANOVA). Values in italics are significant ($p < .05$). }
\begin{tabular}{p{2.5cm}p{1.5cm}p{1.5cm}p{1.5cm}p{1.5cm}p{1.5cm}p{1.5cm}p{1.5cm}p{1.5cm}}
\hline\\
Categories&Animals&Body Parts&Fruits/Veg.&People&Tools&Vehicles&Clothing&\\
Animals&&&&&&&&\\
Body Parts&\it{0.19}&&&&&&&\\
Fruits/Veg.&\it{0.05}&\it{2.5}&&&&&&\\
People&\it{0.36}&663&\it{2.45}&&&&&\\
Tools&\it{0.02}&\it{0.02}&\it{9.9}&\it{0.02}&&&&\\
Vehicles&\it{0.02}&\it{0.05}&\it{11.8}&\it{0.03}&855&&&\\
Clothing&\it{0.02}&\it{0.03}&\it{9.3}&\it{0.02}&973&871&&\\
Part of a House&\it{0.03}&\it{0.06}&\it{2.2}&\it{0.05}&626&728&634&\\
\hline\\
\end{tabular}
\end{table}

\begin{figure}[h!]
\begin{center}
\includegraphics[width=18cm]{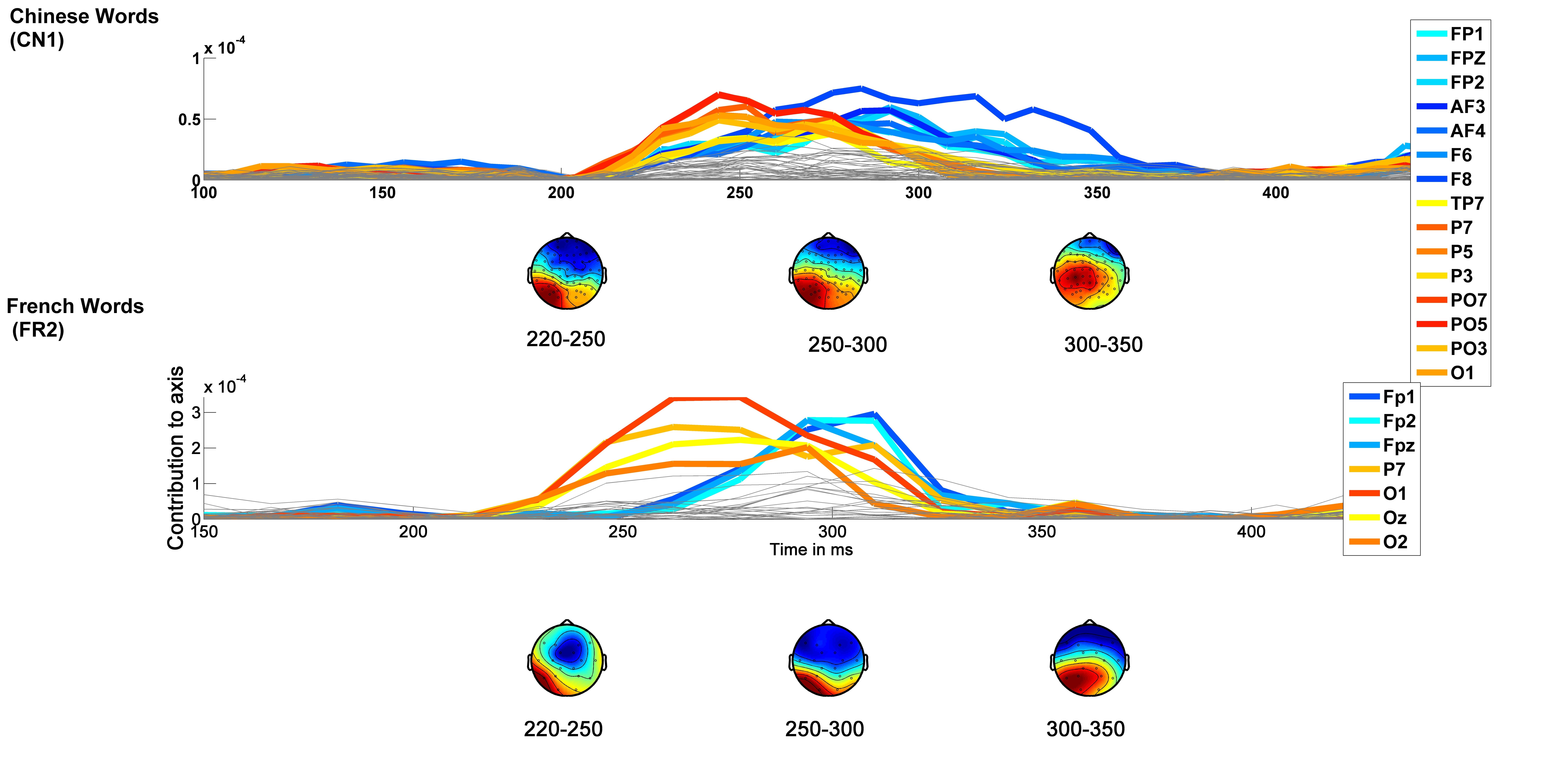}\\
\end{center}
\textbf{\refstepcounter{figure}\label{fig:03} Figure 3.}{ Upper: Contribution of all electrodes to axis CN1 for Chinese words (60 electrodes) and topography of the associated eigenvector in time window $220-350 ms$. Lower: Contribution of all electrodes to axis FR2 for French words (30 electrodes, taken from [Ploux et al., 2012]), and topography of the associated eigenvector in the same time window. }
\end{figure}

\paragraph{Analysis of the electrode x time network underlying axis CN1 (distinction between living and nonliving).} The CA gave the electrode x time network that entered into the calculation of each axis. Figure \ref{fig:03}  represents the time course of the respective contributions of all electrodes to axes CN1 and FR2. These curves clearly show that the distinction between living and nonliving categories occurred between $200$ and $350 ms$ in Chinese, as in French. The main electrodes involved were occipito-parietal, mostly on the left (PO5, PO7, P5, P7, O1, P3, PO3, OZ, O2, TP7, starting at $200 ms$, with a peak around $240 ms$), or frontal, mostly on the right (AF3, F8, AF4, FP2, FP1, FPZ, F6, with a peak around $290 ms$). For French, we found an analogous profile with fewer electrodes (the device we used only had 30 electrodes and the resolution was not as good due to the lower sampling frequency of $500 Hz$). The topography of the eigenvector associated to each of these axes is given in Figure \ref{fig:03} for Chinese and French. It shows an analogy for the time window of interest here ($220-350 ms$). 

\subsubsection{Network Involved in the Processing of Living Categories}
In the French study, the principal axis FR1 for French words indicated differentiation that went from \emph{people} to \emph{fruits/vegetables} for living categories and from \emph{clothing} to \emph{tools} for nonliving categories (Figure \ref{fig:01}), with proximity between \emph{clothing} and \emph{people} on one side, and between \emph{tools} and \emph{fruits/vegetables} on the other. We suggested that this organization revolved essentially around distinctions within the living categories, which were the most highly dispersed and reflected an person-centered reference.
This topology was not given directly by the analysis of the ERP signals obtained here when Chinese words were presented (axis CN2). The differences and similarities between the topologies of Chinese axis CN2 and French axis FR1 will be analyzed in the next section. However, since axis FR1 accounted for the great differences between the living categories, we wanted to find out firstly whether a CA solely on the living categories of Chinese would produce a similar organization to that of French axis FR1, and secondly, how the electrode x time network involved in this living-category organization would position the nonliving categories. To do this, after computing the CA on the living categories only, we projected the nonliving categories onto the resulting axis. Axis 1 of this new analysis was quite similar to axis CN1 calculated from the whole set of categories. On the other hand, axis 2 (denoted CN2b in Figure 1) showed that the organization within the living categories was analogous to French axis FR1 for these same categories. The correlation between the two axes for all eight categories was $corr(CN2b,FR1) = 0.85, p < .01$. Lastly, for Chinese, the projection of the nonliving categories indicated a similar order going from \emph{clothing} to \emph{tools}, although the \emph{clothing} category was farther away from the \emph{people} category than in the results obtained for French on principal axis FR1. 
A repeated-measures Anova on the eight categories for all subjects pooled yielded a significant result ($F(7,112) = 11.5, p < 0.001$). A Duncan post-hoc test indicated three significant differences on this axis: between \emph{people} and each of the other categories ($p(people, other categories) < 0.001$). More generally, all of the living categories differ from each other the coordinates along the two first axes of this CA , and  \emph{people}, \emph{animals}, \emph{body parts } categories differ from each of the other categories ($F(7,128)=22.4, p<.0001$).

The electrode x time network underlying the CN2b organization contains two time windows of interest. The window with the greatest contribution is located between $320$ and $450 ms$. In this window, the electrodes with the greatest contributions themselves are frontal electrodes FPZ, FP1, and F7. The second window of interest is located before the first ($150-230 ms$). The electrodes with the greatest contributions here are (1) occipito-temporal electrodes TP8, O1, Oz, O2, and TP7, and (2) left frontal electrodes AF3, FP1, F1, FC3. For the electrode x time network associated with axis FR1, we found a main window of interest between $350$ and $470 ms$ (frontal electrodes FP1, FP2, FPZ, and F7) and to a lesser extent, an earlier window between $150$ and $250 ms$ (occipito-parietal electrodes P7, P8, O2, and O1, and frontal electrodes FP1, FPZ, and FP2). These two patterns thus contain analogies in terms of both the windows' positions and the location of the electrodes with the greatest contributions.

\subsection{Axis CN2}

Axis CN2 (percentage of inertia explained 16\%) shows an opposition between \emph{people} and each of the other categories, especially \emph{animals}. A repeated-measures Anova ($F(7,112) = 7.8, p < .001$) yielded significant results, with the \emph{people} category differing from each of the others ($p(people, other categories) < .001$), and the \emph{animals} category differing from \emph{people} and \emph{fruits/vegetables} ($p(animals, (people, fruits/vegetables)) < 0.05$.

\begin{figure}[!h]
\includegraphics[width=18cm]{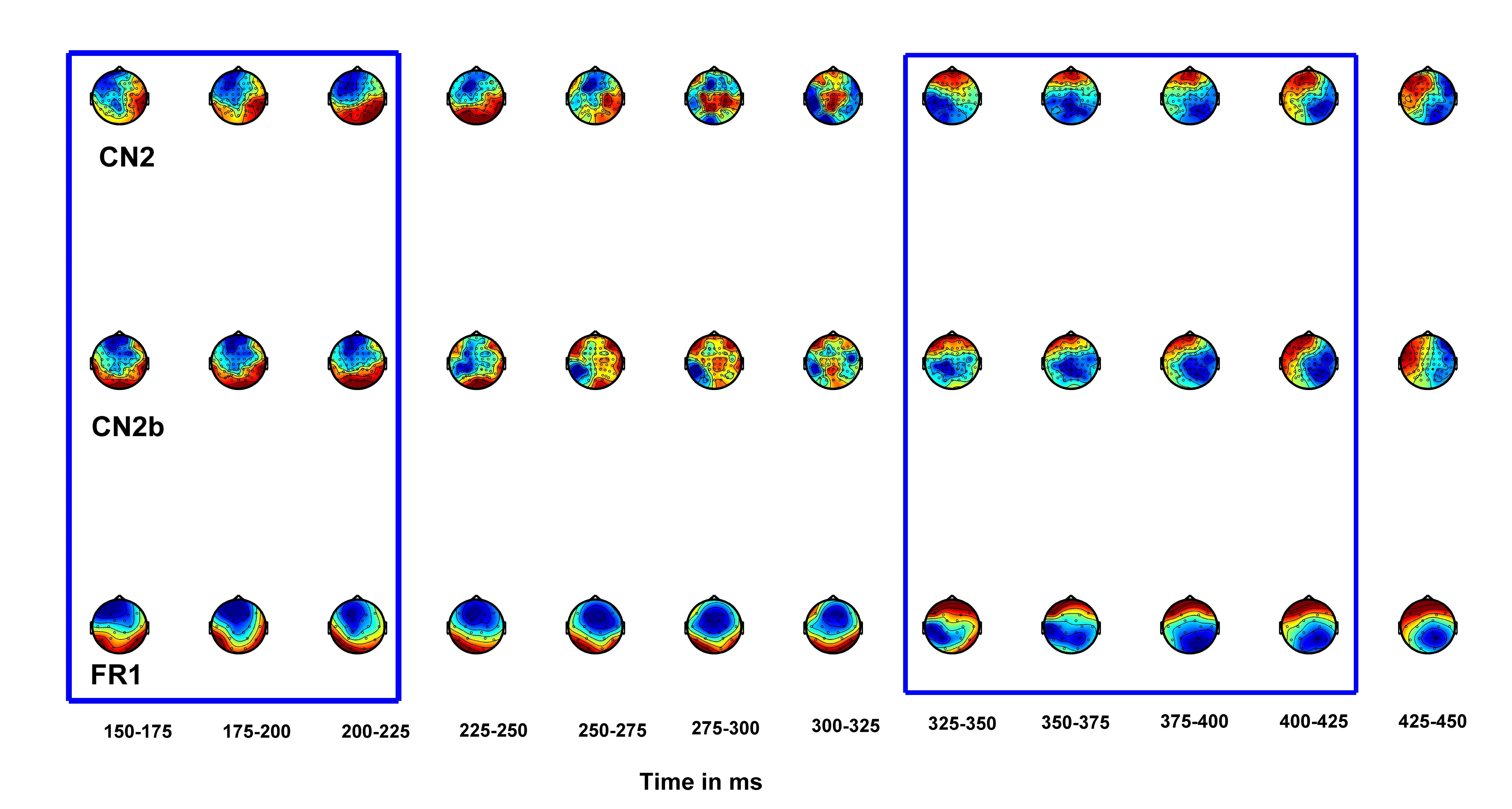}\\
\textbf{\refstepcounter{figure}\label{fig:04} Figure 4.}{ Topography of the eigenvectors associated with each axis (row headings). The rectangles specify the location of the temporal windows with the greatest contributions. }
\end{figure}

The associated electrode x time network contains two windows of interest: the first at $150-230 ms$, with right occipito-parietal electrodes P8, O2, PO6, TP8, and PO8, and left frontal-central electrodes FP1, F3, FC3, FC5, AF3, and FPZ; the second at $330-450 ms$, with frontal electrodes FPZ, AF3, and FP1. 
Figure \ref{fig:04} gives the topographies of the eigenvector associated with each axis. The windows of interest are boxed. 

\section{Comparison with the Organization Given by an Analysis of Word Usage in the Language}

Our last step was to verify whether the organizational similarities and differences found in our analysis of electrophysiological signals would also be found in the output of an analysis of word usage in the language. As in the French study, we used two types of data, based on the fact that they represent distinctive characteristics of semantic links: synonyms or para-synonyms and regular co-occurrences. Synonymy is a relation of the paradigmatic axis (seen as a vertical axis); it is the axis of lexical choices and word substitution (e.g. a word is substituted for another word of similar meaning), whereas co-occurrence is a relation of the syntagmatic axis (seen as a horizontal axis); it is the axis of the sentence and the constituents that co-occur in it (see  \cite{jakobson2002fundamentals}). 

The lexical database of synonyms contains $23,070$ words and $158,755$ synonym pairs\footnote{This database was built using a method -- based on the geometric model of the Semantic Atlas (http://dico.isc.cnrs.fr ) \cite{ploux2009} --  for the creation and the enrichment of synonym and translation dictionaries.  We used as a base of seed synonymy links a subset of the synonymys in the HIT-CIR Tongyici Cilin thesaurus (2005 version, http://ir.hit.edu.cn/demo/ltp/Sharing\_Plan.htm ) \cite{mei1983tongyici}.}.
 We then constructed a semantic space using the geometric model given in \cite{ploux3}. The difference between vector models and our geometric model is that we apply singular value decomposition (SVD) -- CA in this case --  not upon words but upon cliques\footnote{In a previous study, we have shown that this geometric model has better performance than the model LSA \cite{lsabook} in simulating word association \cite{Ji2008}. Cliques of words are fine-grained semantic units. From a mathematical and algorithmic angle, a clique is a set of words all related
to each other by a relationship, synonymy link in the present case.    All cliques containing the words used in the experiment and their synonyms were calculated, for a total of $1456$ words.  Then, the space was created by applying the CA on the matrix  $M$\footnote{$M$ is defined by the formula $M_{i,j} = 1$ if clique $i$ contains term $j$, and $0$ if not.} that contains cliques as  rows and words as columns.} The results indicated a correlation between the first axis of the synonym data and axis CN1 ($corr(CN1,axis1syno) = 0.72, p < .05$), separation of the living and the nonliving ($test-t(923)=5.22, p<0.001$), and more dispersion for living categories ($std=0.22$) than for nonliving ones ($std=0.005$).
The \emph{animals} category is distinct from each other categories  (Anova $F(7,917)=741, p<0.001$, Duncan post-hoc test  $p(animals, other categories)<0.001$)

The same method was applied to a lexical database of co-occurring words generated from a large corpus of texts ($173.4$ million words) in Chinese (NIST-OpenMT06 (see \cite{wang-EtAl:2014:P14-2}) and a United Nation corpus\footnote{LDC2013T06 Info Local CD/DVDL310 ``1993-2007 United Nations Parallel Text" (3 DVDs, Chinese side)}. Inside a window,
which is a sentence in this study, co-occurrences of words were counted and then stored in a database. Co-occurrence links for which the frequency of co-occurrence divided by the product of the word's frequency and its co-occurring word's frequency is less than $0.01$  were removed. 
A second semantic space was then calculated from the words used in the experiment and their co-occurring words, for a total of $1060$ words. The same method as above was applied with cliques of co-occurring words instead of cliques of synonyms. 
The \emph{vehicle} category was removed because, in the corpus we used, the \emph{vehicle} words have a low association with the other words. This induces a huge distance between this category and the others on the built space. For the associated words, we found two correlations, one between axis 1 and axis CN2b ($corr(CN2b,axis1cooc) = 0.8, p < .05$), the other between axis 1 and corpus-data axis CN2 ($corr(CN2,axis1cooc) = 0.78, p < .05$).  The \emph{people} category stood clearly apart from each of the others on this axis, as on axes CN2 and CN2b. A post-hoc test on the coordinates on axis 1 for all words in the experiment, taken one category at a time, showed that the \emph{people} category differed from each of the other living categories and from the \emph{tools} category ($p < .05$). 

As found for French, Chinese exhibited a morphism between the paradigmatic semantic space and the neurophysiological organization of the living/nonliving distinction on one hand, and between the syntagmatic semantic space and the neurophysiological organization of the people/world-objects distinction on the other. 

\section{Discussion}
The present cross-language comparison indicated stability for the following aspects of the languages' lexical semantic organizations. (1) The first aspect is the living/nonliving distinction, which showed up as a main factor for both languages (second factor for French, first factor for Chinese). This distinction is also the first one given by Aristotelian-like conceptual hierarchies (such as WordNet), and is known to play a part in the performance declines of people suffering from semantic dementia \cite{mendez2010interhemispheric}. This not only brings out a similarity in the neurobiological and philosophical importance of this distinction, but also demonstrates the ability of our method to detect it. (2) The second aspect is the greater dispersion of the living categories as compared to the nonliving ones. (3) The third is prototypicality of the \emph{animals} category within the living categories, and with respect to the living/nonliving distinction. (4) The fourth, which occurs within the living categories, is the existence of a gradient that goes from nouns related to \emph{people} to ones related to \emph{fruits and vegetables}, passing through \emph{animals} and \emph{body parts}. A similar gradient opposes clothing-related nouns to tool-related ones. Our electrophysiological analysis indicated stability of the networks at play in each of these processes. Stability was found regardless of the difference between the devices used (60 electrodes for Chinese, 30 for French). It was also observed in the data taken from word usage in the languages, as shown by the results obtained from textual corpora (synonyms and associated words).
These common points raise the question of how the organization shared by these two, very different linguistic systems is linked to the organization of the conceptual system. In this vein, Huth et al. \cite{huth2012continuous} conducted an fMRI experiment in which they applied a data-analysis method to neurophysiological signals. In their experiment, the signals recorded as the subjects watched videos were labelled and segmented by category. The authors obtained semantic spaces whose structures had analogous features to those we obtained for French and Chinese words. In particular, one of the axes (axis 2 in their study) exhibited a gradient going from \emph{people} and people-related characteristics like speech, to objects of the world. Another similarity along one of the axes (axis 4 of their study) was the distance between the living and the nonliving, as in our results, with a preponderance of the \emph{animals} category, the most discriminating category for this distinction. Insofar as the stimuli they presented to the subjects were visual (videos) and nonverbal, we can contend that the stability of these aspects -- found via both a language modality and a perceptual modality, and with various tasks -- is supported by a common fundamental cognitive organization. 
We would also like to emphasize that while the task in the present study required a categorization that may have underlain the observed differentiation,
the living/nonliving differentiation was not the greater one detected in the signal analysis for French words stimuli (axis 2). In addition, this differentiation also appears in the very first axes (axis 4) in  \cite{huth2012continuous} study containing no categorization task. Furthermore, one can observe that the results obtained are more precise and more complex than a simple binary categorization since the \emph{animals} category,  as said above, stands well apart from the others 
(axis 1 for Chinese words, axis 2 for French words, axis 4 in \cite{huth2012continuous} visual study).  
This could be interpreted as a more clear-cut, more prototypical attribution of aliveness to \emph{animals} as opposed to \emph{fruits/vegetables}, \emph{body parts}, or \emph{people} (this last category's attributes being at least as social as biological).

Lastly, as we mentioned in the introduction, a debate is current underway between advocates of an abstract, amodal semantic system, and advocates of an "embodied" approach to cognition in which the processing of word meaning calls upon shared sensorimotor areas activated during the perception of objects or actions being carried out. Some more recent studies \cite{jouen2015beyond} tend to show that this shared system recruits other systems too, i.e., not only the above sensorimotor areas, but also systems such as episodic memory, reasoning, and so on. Our results support the idea that the structure of the lexicon is not based solely on the sensorimotor features of named objects. This follows from the fact that the living/nonliving distinction -- and the fact that \emph{fruits and vegetables}, for example, are "more" living than \emph{tools} -- cannot be accounted for in terms of the sole utilization of the sensorimotor modalities, but requires access to the morphogenesis of the entity (\emph{vegetables and fruits} grow and develop like all living things, which cannot be said of \emph{tools}, \emph{vehicles}, etc. which are made and assembled by human beings). Likewise, the distinction between \emph{people} and entities of the world (axes CN2, CN2b, and FR1) shows that the processes at play in certain semantic distinctions are anchored in a manmade referential not found in the conceptual hierarchies of the Aristotelian type, nor in their alternatives based on the notion of prototypicality \cite{rosch1973natural}. Moreover, because this referential clearly distinguishes \emph{body parts} from the \emph{people} category, we can contend that the organization of this axis is not rooted in perceptual or motor features but in other characteristics anchored in the recognition of congeners (i.e., for a human subject, nouns pertaining to people), who share the same development and potentially the same range of actions, perceptions, social interactions, and worldly intelligence. 

This being said, the results obtained for both languages (French and Chinese) indicate not only qualitative differences but also differences in degree. Among the qualitative differences, we can see that axis CN2 has no French equivalent and reflects an organization gradient that is difficult to interpret from the semantic standpoint. An investigation of how the two systems differ linguistically could provide some indications, since French has an alphabet while Chinese has ideograms and an extended system of classifiers that French doesn't have. Ideograms can contain elements that classify the word, for instance, in the \emph{people} category (\begin{chinois}人\end{chinois}, r\'en), the \emph{birds} category, or the \emph{things} category (\begin{chinois}子\end{chinois},z\`\i), etc.  These elements cannot, however, establish an ordered gradient (like the one we found here:  \emph{animal} nouns on top, then \emph{fruit and vegetable} nouns, then \emph{people} nouns). Similarly, the Chinese classifiers -- which are items of measurement or designations associated to certain types of words and may therefore have a grouping effect on the words to which they pertain -- can themselves pertain to different categories. For example, (\begin{chinois}个\end{chinois}, g\`e) is one of the classifiers of \emph{people} but also of \emph{body parts}, \emph{fruits}, and objects like refrigerators. Another example is the classifier (\begin{chinois}只\end{chinois}, zh\v{\i}) for nouns referring to animals: it is also a classifier of \emph{body parts} that come in pairs (eyes, hands, etc.). Classifiers, then, are not useful for understanding the organization of the axes as a whole, nor of axis CN2 in particular. 
Lastly, we can propose the hypothesis (which remains to be validated) that axis CN2, like axis CN2b (and to a lesser extent axis FR1), distinguishes the people category from the set of all other categories, and that it represents whatever characterizes this distinction, either from the standpoint of the set of all living categories (axis CN2b) or from the standpoint of the set of all categories (axis CN2). Accordingly, for Chinese axis CN1 (living vs. nonliving), \emph{fruits/vegetables} has a coordinate close to those of nonliving categories, closer than that obtained for French axis FR2. This lends support to the hypothesis that the weight, volume, and graspableness features common to \emph{fruits/vegetables} and manmade objects, and the functional features common to \emph{body parts} and those same manmade objects (e.g. tools or vehicles), tend to group them together. The CA, then, could point to a second organization and major distinction: the opposition between \emph{people} and \emph{animals}. 
Concerning the differences in degree -- such as the slight differences between the locations of the categories on axis CN1 and axis FR2, or axis CN2b, or the differences in the order of the inertia values of the axes (the fact that the living/nonliving distinction was first in Chinese and second in French) -- we can bring several possible causes to bear. The first is methodological: the use of different devices may have had a slight impact. The second is more fundamental: there may be variations due to the effects of proximities and distinctions that occur whenever the words in question form a network, a system. Such differences could, for example, be related to the effects of polysemy and context-dependent variations in word meaning. Clearly, the vast majority of words have several meanings, and these different meanings are not necessarily shared across languages. A case in point is the word \emph{bureau} in French, which can refer to a work table or a room where one works. The link between these two meanings, created by the use of the same word for both, does not exist in Chinese (nor in English for that matter). When one studies these spaces based on language-production data \cite{ploux3}, polysemy has different distortion effects on the topology of the semantic spaces in each linguistic system, and in a similar way, could lead to analogous differences in the spaces constructed from neurophysiological signals measured during the semantic processing of words.

\newpage

\section*{Data Sharing}
Result files and source code are available upon request to the corresponding author.

\section*{Disclosure/Conflict-of-Interest Statement}

The authors declare that the research was conducted in the absence of any commercial or financial relationships that could be construed as a potential conflict of interest.

\section*{Author Contributions}

S.P., Z.H., B-L.L., conceived and supervised  the project. Z-F.Z. and R.W. developed  the stimuli. R.W. and Y.X.  designed the experiment and acquired the data. S.P. and A.C. created the comparison model between the two languages. S.P. wrote the article. All authors discussed the results and implications and commented on the manuscript at all stages.

\section*{Acknowledgments}
This study received financial support from the Cai YuanPei Program of the French Ministry of Foreign Affairs.

\newpage
\bibliographystyle{apacite}


\end{document}